\documentclass{evn2002}
\usepackage{graphicx}
\newcommand{\cthead}[1]{\multicolumn{1}{c}{#1}}
\newcommand{\kss}{km~s$^{-1}$ }
\newcommand{\ks}{km~s$^{-1}$}

\begin{document}
\title{EVN observations of 6.7 GHz methanol masers from Medicina
survey}

\author{M.A. Voronkov\inst{1}, V.I. Slysh\inst{1}, F. Palagi\inst{2}
          \and G. Tofani\inst{2}}

   \institute{Astro Space Center, Profsouznaya st. 84/32, 117997 Moscow, Russia
         \and
Osservatorio Astrofisico di Arcetri, Largo E. Fermi 5, 50125 Firenze, Italy}

   \abstract{
We report VLBI observations of methanol masers in the brightest
$5_1-6_0$~A$^+$ transition at 6.7 GHz in NGC 281W, 18151$-$1208 and
19388$+$2357. Using the fringe rate method absolute positions were obtained
for all observed sources. A linear ordered structure with a velocity
gradient was revealed in NGC 281W. Under assumption that such structure
is an edge-on Keplerian disk around the central object with a mass of
30$M_\odot$ located at a distance of 3.5 kpc from the Sun, we estimated
that methanol masers are situated at the distance about 400 a.u. from the
center of the disk. A second epoch of observations was reported
for L1206, GL2789 and 20062$+$3550. The upper limits on the relative motions of
maser spots are estimated to be 4.7~\kss and 28~\kss for L1206
and GL2789 respectively.
   }

\authorrunning{M.A. Voronkov et al.}
   \maketitle
%

\section{Introduction}
The $5_1-6_0$~A$^+$ methanol transition at 6.7~GHz produces the
brightest known methanol masers. Many recent interferometric studies of such
masers reveal geometrically ordered structures formed by maser spots
sometimes with velocity gradients (Norris et al. \cite{nor93},
Phillips et al. \cite{phi98}, Minier et al. \cite{min00}). Such linear
structures can
be explained in the model of the rotating Keplerian disk seen edge-on.
The goal of this work is to map using EVN several methanol masers
discovered in the Medicina survey (Slysh et al. \cite{sly99}) and to
search for geometrically ordered structures in these sources. 

\section{Results and Discussion}
The EVN observations were carried out in 1998 and 2000. In total, images of
6 sources were constructed, with 3 of them being imaged in both epochs of
observations. Analysis of these images yields relative positions and flux
densities of individual maser spots which were shown in Table~\ref{features}.
During the postpocessing absolute positions of the reference features
were determined using the fringe rate method. Two sources, L1206 and GL2789,
were described in detail earlier by Voronkov \& Slysh (\cite{vor01}) and
Val'tts et al. (\cite{val02}). Among the sources observed in both sessions
the two masers are the only sources which consist of multiple components and
so may be used to study relative motions of maser spots. The upper limits
on this motions are estimated to be 4.7~\kss and 28~\kss for L1206 (1~kpc)
and GL2789 (6~kpc)  respectively.
For the sources 19388$+$2357 and
20062$+$3550 we detected only one spectral feature in the correlated spectrum.
The lower limits on the brightness temperature are $5.9\times10^{11}$~K and
$1.3\times10^{10}$~K for 19388$+$2357 and 20062$+$3550 respectively.

\begin{table}[!t]
\caption{Relative positions and flux densities of individual maser spots.
Typical positional accuracy is about 2 mas.}
\label{features}
\begin{center}
\footnotesize
\begin{tabular}{lllrrr}
\cthead{Source} & &\cthead{V$_{LSR}$}& \cthead{$\Delta\alpha$$^a$} &\cthead{$\Delta\delta$} &
\cthead{S}\\
& &\cthead{(\ks)}&\cthead{(mas)}&\cthead{(mas)}&\cthead{(Jy)}\\
\hline
\multicolumn{6}{c}{\it 1998 year}\\
20062$+$3550& & $-$2.62 &&& 2.2 \\
\rule{0pt}{11pt}%
GL2789      &A& $-$43.37& $-$9 & 20.4 & 2.0 \\
            &B& $-$42.5 & 0 & 0       & 2.5 \\
	    &C& $-$40.74& $-$2.5& $-$68.9& 2.0 \\
	    &D& $-$40.39& 9.2 & $-$78 & 2.2  \\
\rule{0pt}{11pt}%
L1206       &A& $-$11.0 &  0  & 0     & 17.0 \\
            &B& $-$10.2 &  $-$93.7 &  35.9  & 1.0 \\
	    & &         &  $-$85.7 &  28.0  & 1.0 \\
\multicolumn{6}{c}{\it 2000 year}\\
NGC 281W    &A& $-$29.35&  0  & 0     & 6.3  \\
            &B& $-$29.0 & 4.7 & 0.2   & 7.8  \\
\rule{0pt}{11pt}%
18151$-$1208&A& $+$27.65& 0   & 0     & 2.8  \\
            &B& $+$28.1 &$-$4.2& $-$3.4&5.9 \\
	    & &         &$-$2.8& $-$2.2&1.7  \\
\rule{0pt}{11pt}%
19388$+$2357& & $+$38.21&      &      & 18.0 \\
\rule{0pt}{11pt}%
20062$+$3550& & $-$2.6 &&& 1.2 \\
\rule{0pt}{11pt}%
GL2789      &A& $-$43.55 & $-$7 & 20 & 2.5  \\
            &B& $-$42.5  & 0 & 0 & 1.2 \\
	    &C& $-$40.82 & $-$2.1 & $-$69 & 1.7 \\
	    &D& $-$40.48 & 10& $-$78 & 3.0 \\
\rule{0pt}{11pt}%
L1206$^b$   &A& $-$11.0 &  0 & 0     & 1.7 \\
            & &         & $-$0.2 & $-$6.0 & 2.1 \\
            &B& $-$10.2 & $-$84.1 & 25.6  & 0.3 \\
\hline
\end{tabular}
\end{center}
\footnotesize
\noindent
$^a\hphantom{b}$~-- Taking into consideration of the $\cos\delta$ factor.\\
\noindent
$^b\hphantom{a}$~-- The maps of the reference feature are considerably different for two
epochs, so the error of relative positions may be high.
\vskip -7mm
\end{table}

\begin{figure}[!t]
\centering
\includegraphics[width=0.895\linewidth]{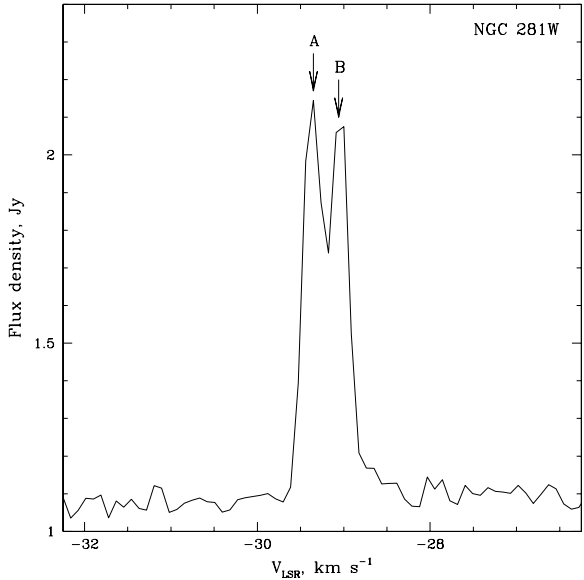}
\caption{The correlated spectrum of the methanol maser in NGC~281W on
the short baseline Effelsberg~-- Jodrell Bank.}
\label{ngc281wsp}
\end{figure}

\begin{figure}[!t]
\centering
\includegraphics[width=\linewidth]{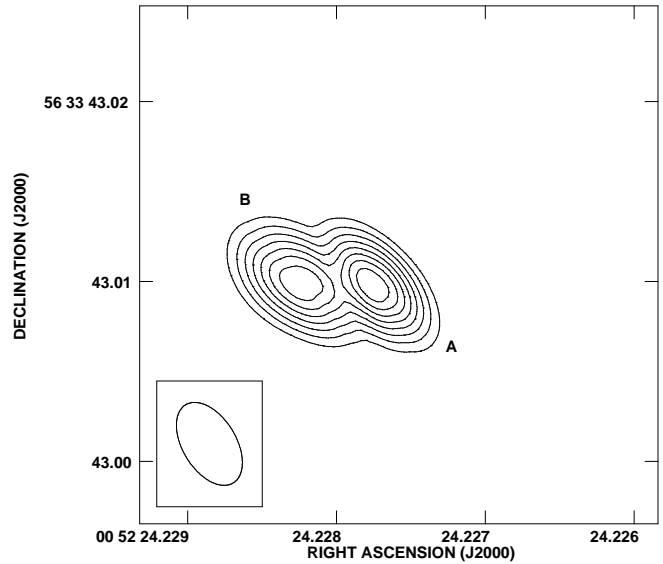}
\caption{The map of the methanol maser in NGC~281W. Contours are
0.6$\times$(2, 3, 4, 5, 6, 7, 8, 9) Jy~beam$^{-1}$.}
\label{ngc281wmap}
\end{figure}

In addition to GL2789 and L1206, where the spatially ordered structures
with a velocity gradient were revealed in the previous studies cited above,
such a structure was found in the maser NGC~281W. The spectrum and the
map of this source are shown in Fig.~\ref{ngc281wsp} and
Fig.~\ref{ngc281wmap}. The lower limit on the brightness temperature is
$1.9\times10^{10}$~K being almost the same for both spectral features. 
The map of each spectral channel is a point-like
source. The position of this point source shifts from the spot A to the
spot B in Fig.~\ref{ngc281wmap} while moving the selected spectral channel
across the spectrum from the feature A to the feature B in
Fig.~\ref{ngc281wsp}. Such a gradient can be explained in the model
where masers are formed in the Keplerian disk seen edge-on
(Minier \cite{min00}). In this case the observed gradient
$\frac {d\;V}{d\;x}$ of the radial velocity is
\begin{equation}
\frac {d\;V}{d\;x}=30\;\mbox{km}\;\mbox{s}^{-1}\;\mbox{a.u.}^{-1}
\sqrt{\frac{M/M_\odot}{R_{\mbox{\scriptsize a.u.}}^3}}\mbox{,}
\label{gradient}
\end{equation}
where $M$ is the mass of the central object (protostar),
$R_{\mbox{\scriptsize a.u.}}$
is the distance between the maser and the protostar (in a.u.).
Assuming that the mass of the central object is about 30$M_\odot$ and
the distance from the Sun is about 3.5~kpc we estimate
$R_{\mbox{\scriptsize a.u.}}$ to be about 400 a.u.

The spectrum of the source 18151$-$1208 consists of two features A and B
(see Table~\ref{features}). The maser spot corresponding to the most
intense feature B was resolved into two spots at almost the same velocity.
The maps of individual spectral channels show no such behavior as it does
in the case of NGC~281W. This source shows fringes on the longest baseline
Effelsberg$-$Hartebeesthoek (about 8000 km) which is probably unusual for
the 6.7~GHz methanol masers. The lower limit on the brightness temperature
is $1.4\times10^{10}$~K for this source.

\section{Conclusions}
The geometrically ordered structure with a velocity gradient was
revealed in the 6.7 GHz methanol maser in star-forming region NGC~281W.
Being interpreted as an edge-on Keplerian disk around a protostar with
the mass about 30$M_\odot$ located at the distance of 3.5~kpc from the Sun
such a structure implies the model where methanol masers are situated at the
distance of 400 a.u. from the central object. This result is similar to
that obtained for L1206 and GL2789. The brightness temperature for the
most compact maser 19388$+$2357 exceeds $5.9\times10^{11}$~K.
The source 18151$-$1208 shows fringes on the 
Effelsberg$-$Hartebeesthoek (about 8000 km) baseline. The upper limits on
the relative motions of
maser spots are estimated to be 4.7~\kss and 28~\kss for L1206
and GL2789 respectively.
\begin{acknowledgements}
The European VLBI Network is a joint facility of European, Chinese and
other radio astronomy institutes funded by their national research councils.
This project was partially supported by the INTAS grant no. 97-1451.
MAV and VIS were supported by the RFBR grant no. 01-02-16902. MAV was also
supported by RFBR grant no. 02-02-06800 
and the Radio Astronomy Research and Education Center.
\end{acknowledgements}

\end{document}